\documentclass[letter]{aa}  

\usepackage{graphicx}
\usepackage{txfonts}
\usepackage{lipsum}

\usepackage{subcaption}         
\usepackage{lscape}             
\usepackage{placeins}           

\defcitealias{VS23}{VS2023}
\defcitealias{deJager88}{dJ88}
\defcitealias{hainich}{H14}
\usepackage{hyperref}

\begin{document}

   \title{Wolf-Rayet formation through $L/M$-dependent superwinds}

   \subtitle{}

%

   \author{Aisling McKeown\inst{1,} \inst{2}\fnmsep\thanks{Corresponding author: amckeown58@qub.ac.uk }\and Ciar\'{a}n Furey \inst{1,} \inst{2} \and Gautham N. Sabhahit \inst{1} \and Jorick S. Vink \inst{1}}

   \institute{ Armagh Observatory, College Hill, Armagh BT61 9DG, UK
   \and
   School of Maths and Physics, 
Queen’s University Belfast, University Road, Belfast, BT7 1NN, UK}

   \date{}

 
  \abstract
   {Mass-loss in red supergiants (RSGs) is poorly understood and stellar evolution models rely on empirical prescriptions. In recent years, new superwind mass-loss ideas -- including an $L/M$ dependence -- have been proposed, unlike traditional RSG wind recipes which are formulated on a $L$ dependence alone. We investigate how contrasting RSG mass-loss prescriptions affect the late-time evolution of massive stars from their RSG phase onwards.  We used MESA to create a series of stellar evolution models with masses ranging from $10\,{\rm M}_\odot - 40\,{\rm M}_\odot$, using the RSG \cite{VS23} $L/M$ recipe and traditional mass-loss prescriptions. Independent of the treatment of the stellar envelope, we find for a 25$\,{\rm M}_\odot$ star that the $L/M$ recipe reaches an order of magnitude higher mass-loss rate than traditional recipes due to its inverse mass dependence. As the RSG luminosity is set by the core mass and becomes largely insensitive to the current stellar mass, this creates a positive feedback loop driving runaway envelope stripping. Consequently, stars with an initial mass $M_{\rm init} \sim \,20\,{\rm M}_\odot$ are able to strip their hydrogen envelope and form classical Wolf-Rayet (WR) stars. We draw this conclusion not solely on the basis of their HRD location, but also on their predicted emission-line spectrum that we compute with the PoWR stellar atmosphere code. Stars with $M_{\rm init}<20\,{\rm M}_\odot$ do not fully strip, meaning they expand after core helium burning. In Appendix A we present a generic analytical model showing that the tendency for runaway stripping increases rapidly with the strength of the inverse mass-loss scaling with mass. Our models demonstrate a viable single-star pathway into the yellow supergiant and WR regimes. The amounts of full and partial stripping have a direct effect on supernova (SN) progenitors, including partially stripped type IIb SNe.}


   \maketitle
\nolinenumbers

\section{Introduction}
Massive stars ($M\geq8\,{\rm M}_\odot$; \citealp{Poelarends}) experience complex evolutionary pathways driven by nuclear burning, internal mixing, and mass-loss.
Once core hydrogen (H) burning has finished and core helium (He) burning begins, stars with sufficiently extended envelopes and effective temperatures below 5000\,K are classified as red supergiants (RSGs).

Mass loss during the RSG phase plays a critical role in determining the final fate of the star. 
Despite its importance, the physical mechanism driving mass-loss in RSGs remains poorly understood \citep[for a review, see e.g.][]{Decin}. Therefore, current stellar evolution models rely on empirical or semi-empirical prescriptions to estimate their wind strengths. 

The analysis of RSG mass-loss prescriptions is not new. Early stellar evolution models often adopted a Reimers-type law \citep{Reimers75}, whereas more recent work has typically relied on the prescription by \citet[hereafter \citetalias{deJager88}]{deJager88}. However, although this widely used prescription incorporates a dependence on luminosity and effective temperature, it does not explicitly account for stellar mass. 


A range of mechanisms have been proposed to enhance mass-loss during the RSG phase, including pulsation-driven superwinds \citep[][]{Herger,Yoon_2010}, radiative driving near the Eddington limit \citep[][]{2016MNRAS.455..112G,VS23}, and turbulence-enhanced outflows \citep[][]{Kee,Fuller_2024}. While these mechanisms differ in their physical origin, they share the common outcome of increasing mass-loss in evolved massive stars.

Recently, \citet[hereafter \citetalias{VS23}]{VS23} derived an empirical mass-loss rate ($\dot{M}$) prescription for RSGs based on the data of \cite{Yang2023} which directly relates $\dot{M}$ to the stellar luminosity-to-mass ratio; i.e. $\dot{M} \propto L/M$.
This formulation is more physically motivated than traditional empirical laws because it incorporates a strong inverse dependence on the current stellar mass, reflecting the increasing susceptibility of the stellar envelope to wind driving as the star evolves. \citetalias{VS23} further demonstrated that this prescription naturally matches the empirical upper luminosity limit for RSGs, without requiring additional, potentially artificial, energy transport mechanisms in the super‑adiabatic layers, such as those invoked through MESA’s MLT++ \footnote{MLT++ is an extension of standard mixing-length theory that artificially reduces superadiabaticity in radiation-dominated envelopes to improve numerical convergence.} parametrisation.

In this work, we compare the \citetalias{deJager88} and \citetalias{VS23} RSG $\dot{M}$ prescriptions for massive stars at Large Magellanic Cloud (LMC) metallicity and assess their impact on stellar evolution. In Sect. \ref{sec:methods}, we describe the details of our stellar evolution calculations and the implementation of the \citetalias{VS23} mass-loss prescription, and in Sect.~\ref{sec:results} we present its effect on RSG evolution for a range of initial masses.
\section{Methods}\label{sec:methods}

\subsection{Initial Masses and Metallicity}

We used the Modules for Experiments in Stellar Astrophysics (MESA) stellar evolution code version r24.08.1 \citep{Paxton2011,Paxton2013,Paxton2015,Paxton2018,Paxton2019} to evolve stars with initial masses between 15$\,{\rm M}_\odot$ and 40$\,{\rm M}_\odot$. The metallicity ($Z$) is the abundance of all elements which are not H and He. We evolved our models at LMC metallicity \citep[$0.5\,{\rm Z}_\odot$;][]{Clayton_2025} which corresponds to a metal mass fraction $Z=0.01$. We evolved non-rotating models from the pre-main sequence until post core He burning. 
Further details about the models are presented in Appendix~\ref{app:MESA_models}.
\subsection{Mass-Loss}
We ran our evolution models with two contrasting RSG mass-loss recipes\footnote{We consider stars to be in the RSG phase when $T_{\mathrm{eff}}< 5000\,$K.}. The traditional \citetalias{deJager88} prescription parametrises $\dot{M}$ with a luminosity and temperature scaling given by:
\begin{equation}
\log \left( \frac{\dot{M}_{\rm dJ88}}{ {\rm M}_\odot \, {\rm yr}^{-1} } \right)
= -8.158 + 1.769 \log \left( \frac{L}{{\rm L}_\odot} \right) 
- 1.676 \log \left( \frac{T_{\rm eff}}{\mathrm{K}} \right),
\label{eqn:dj88}
\end{equation}
where $T_{\mathrm{eff}}$ is the effective temperature. Notably, this prescription contains no explicit dependence on the current stellar mass, $M$.

The other was the \citetalias{VS23} prescription, which does incorporate an inverse $M$ scaling to account for the effect of gravity: as $M$ decreases, so does the surface gravity, and consequently $\dot{M}$ increases. Unlike in O stars, where luminosity is coupled to the current mass, the luminosity of an RSG is determined primarily by the He core mass set at the end of the main sequence (MS), 
and therefore decouples from the current total mass. This introduces a feedback effect in which decreasing mass leads to progressively higher $\dot{M}$. The \citetalias{VS23} prescription defines a kink between a low $L/M$ regime and a high $L/M$ regime, given by:

\begin{equation}
   \log \left (\frac{\dot{M}_\mathrm{lowL/M}}{{\rm M}_\odot \, {\rm yr}^{-1}}\right) =-8 + 0.7\log\left(\frac{L}{{\rm L}_\odot}\right) -0.7\log\left(\frac{M}{{\rm M}_\odot}\right) ,
\label{eqn:vs_mdot_lowL}
\end{equation}

\begin{equation}
  \log \left (\frac{\dot{M}_\mathrm{highL/M}}{{\rm M}_\odot \, {\rm yr}^{-1}}\right)  = -24 + 4.8\log\left(\frac{L}{{\rm L}_\odot}\right) -3.99\log\left(\frac{M}{{\rm M}_\odot}\right) .
\label{eqn:vs_mdot_highL}
\end{equation} 

The sensitivity of the total mass lost and the tendency for a mass-loss runaway, both of which depend on the scaling of $\dot{M}$ with mass (zero for the~\citetalias{deJager88} prescription and $-3.99$ for the~\citetalias{VS23} prescription), are explored via a simple analytical model in Appendix~\ref{app:analytical_model}. We present the wind prescriptions used beyond the RSG winds in Appendix~\ref{app:Other Winds}.

\section{Results and Discussion}\label{sec:results}
\subsection{Treatment of Superadiabaticity} \label{subsec:MLT++ Treatment}

\begin{figure}
    \centering
    \includegraphics[width=1.0\linewidth]{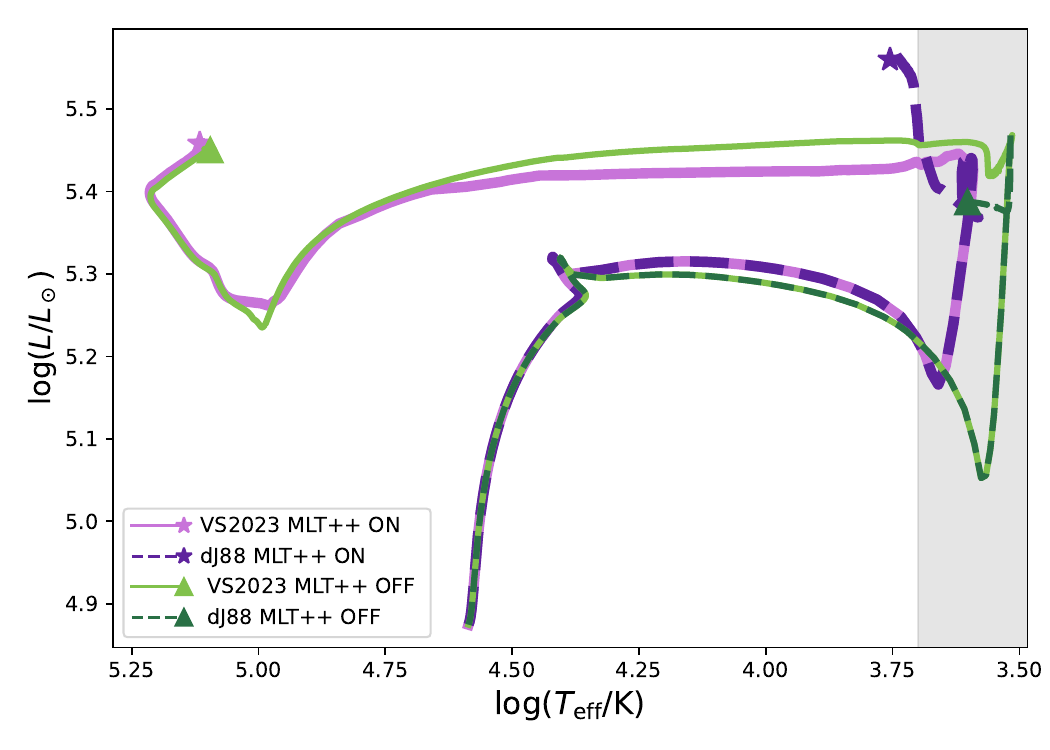}
    \caption{HRD for a 25 $\,{\rm M}_\odot$ star with MLT++ on and off using the \citetalias{deJager88} prescription (dashed lines) and \citetalias{VS23} prescription (solid lines). The symbols $\blacktriangle$ and $\star$ denote the endpoints.
    The point at which the MESA models switch to the RSG recipes is shown by the grey background. "Endpoint" refers to the termination of the model run, which is caused by either core He exhaustion or numerical limitations.} 
    \label{fig:MLT++_HRD}
\end{figure}

We find, for a 25$\,{\rm M}_\odot$ star, that the traditional \citetalias{deJager88} prescription evolves to the RSG phase and remains there until the end of its evolution 
as seen in Fig.~\ref{fig:MLT++_HRD}. This behaviour is a consequence of the strong $T_{\mathrm{eff}}$ dependence in Eqn.~\ref{eqn:dj88}. While the \citetalias{deJager88} prescription yields higher $\dot{M}$ at lower $T_{\mathrm{eff}}$, the resulting total mass-loss in the RSG regime is insufficient to fully strip the H envelope. Consequently, the star remains in the RSG phase until core collapse. 

The \citetalias{VS23} recipe forms a RSG before evolving blue-wards to form a Wolf-Rayet (WR) star. The enhanced mass-loss allows the 25$\,{\rm M}_\odot$ star to strip its envelope and form a WR star; this (as well as the behaviour of the \citetalias{deJager88} models described above) occurs regardless of the treatment of energy transport in the super-Eddington, superadiabatic envelope layers of the stars, i.e., whether standard mixing length theory (MLT) or MLT++ is used in MESA. We find that for the \citetalias{VS23} recipe there is no change in the physical endpoint with MLT++ on or off. Therefore, we adopt MLT++ on for all subsequent models presented in this letter for computational efficiency.

While both \citetalias{VS23} models and the \citetalias{deJager88}(MLT++ ON) model successfully reach the endpoint of core He exhaustion (i.e. where $Y_{\rm{c}} = 0.00$), the \citetalias{deJager88}(MLT++ OFF) model terminates with a central He fraction $Y_{\rm{c}}= 0.221$. This early termination occurs because the code hits its minimum timestep limit; severe envelope inflation prevents the solver from converging on a stable hydrostatic solution. Because this specific endpoint is dictated by numerical limitations rather than nuclear fuel exhaustion, care must be taken when comparing these tracks, as the \citetalias{deJager88}(MLT++ OFF) model does not represent a completed He burning phase.


\subsection{How mass-loss varies between the two recipes}\label{subsect:deltamdot}

\begin{figure}
    \centering
    \includegraphics[width=1.0\linewidth, height = 0.76 \linewidth]{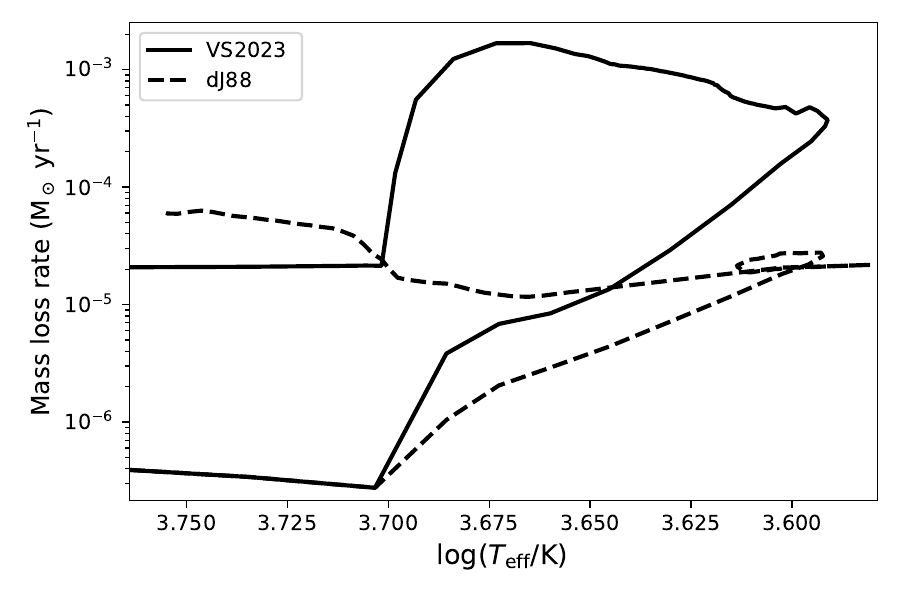}
    \caption{
    Mass-loss rate as a function of $T_{\rm eff}$ during the RSG phase of a 25$\,{\rm M}_\odot$ star, comparing the \citetalias{VS23} and \citetalias{deJager88} recipes.}
    \label{fig:rates}
\end{figure}

For a $25\,{\rm M}_\odot$ star, we find that the \citetalias{deJager88} and \citetalias{VS23} mass-loss prescriptions diverge at the onset of the RSG phase (where $\log(T_\mathrm{eff}/ \mathrm{K}) \approx 3.7$). This is demonstrated in Fig.~\ref{fig:rates}, which shows that the \citetalias{VS23} recipe peaks with $\dot{M} \approx 10^{-3}\ \mathrm{M_\odot \ yr^{-1}}$. The \citetalias{deJager88} has a considerably lower $\dot{M}$ than the \citetalias{VS23} recipe, as it reaches a maximum $\dot{M} \approx 10^{-5}\ \mathrm{M_\odot \ yr^{-1}}$. As shown in Appendix~\ref{app:analytical_model}, the $L/M$ dependence introduces a positive mass-loss feedback. For the \citetalias{VS23} scaling with $f_\mathrm{m} = 3.99$, this can trigger runaway envelope stripping, whereas the absence of a mass dependence in the \citetalias{deJager88} recipe limits the total mass lost in the RSG phase. Consequently, the \citetalias{deJager88} model remains within the RSG region of the HRD (Fig.~\ref{fig:MLT++_HRD}) because $L$ and $T_{\mathrm{eff}}$ do not significantly change during this phase of evolution. 
In contrast, the $L/M$ dependent mass-loss enhancement in the \citetalias{VS23} recipe is sufficient for the star to strip its outer envelope and form a WR star.

\subsection{The Effect of RSG superwinds on post-RSG evolution} \label{subsect:effect}
\begin{figure}
    \centering
    \includegraphics[width=1.0\linewidth]{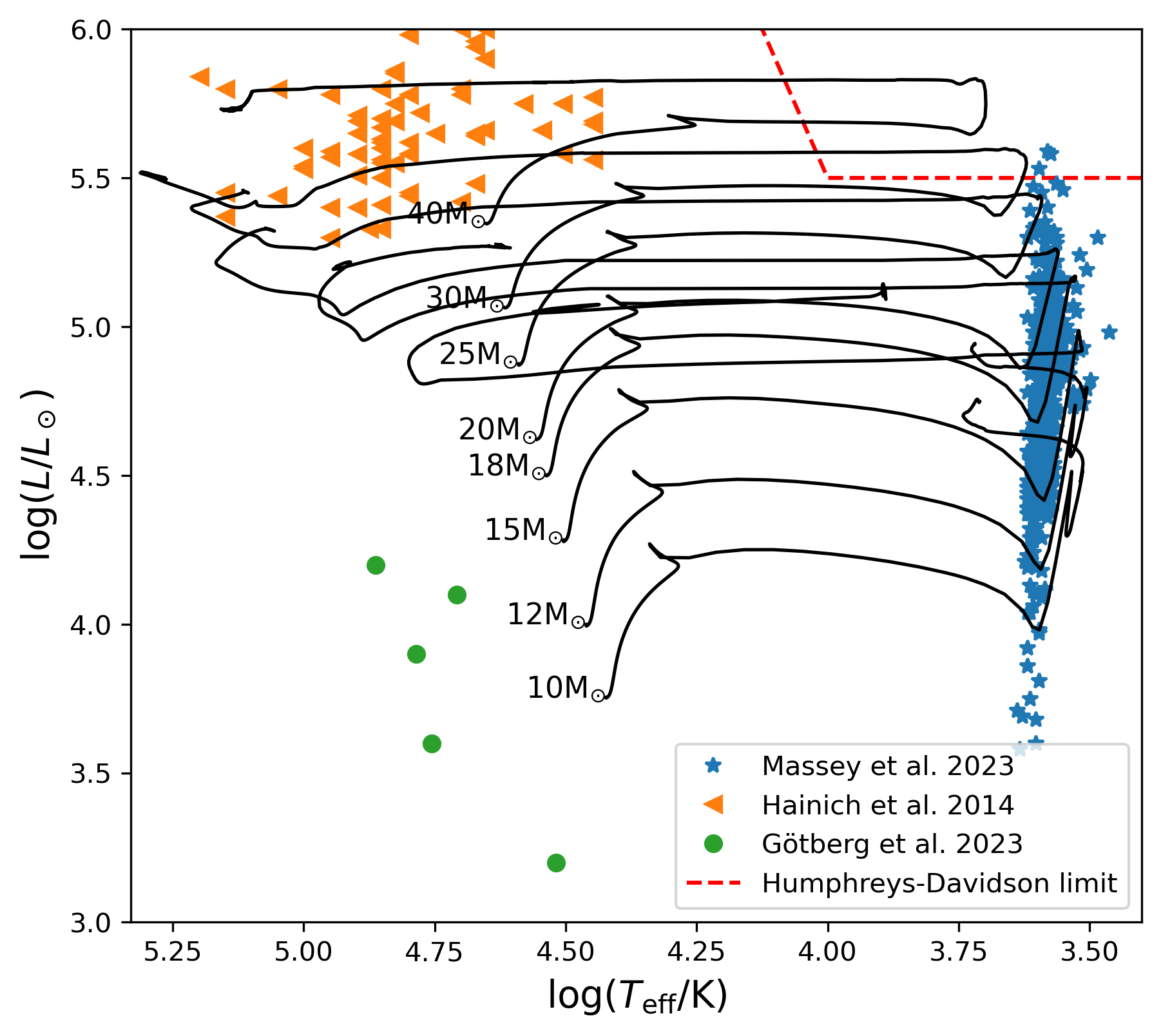}
    \caption{HRD showing the evolution of our stars using the \citetalias{VS23} recipe, with observational data for RSGs \citep{Massey23}, WN stars \citepalias{hainich} and intermediate mass stripped He stars \citep{Gotberg}. The Humphreys-Davidson (HD) limit is indicated by the red dashed line \citep{Davies}. Note that the quoted temperatures of the \citetalias{hainich} sample are defined at Rosseland optical depth $\tau_{\rm Ross} = 20$, meaning these points should be shifted by approximately 0.2 dex lower in $\log(T_{\rm eff}/{\rm K})$ to be compared with $T_{\rm eff}$ reported by MESA.    
}

\label{fig:fullgrid}
\end{figure}

Fig.\ref{fig:fullgrid} demonstrates that the most luminous RSG models reach luminosities comparable to the observed upper end of the RSG population from \cite{Massey23}. These models subsequently evolve bluewards as mass-loss removes a significant fraction of the H envelope. The resulting post-RSG models occupy a region of the HRD consistent with the WR stars identified by \citet[hereafter \citetalias{hainich}]{hainich}. This suggests that, using the \citetalias{VS23} enhanced mass-loss prescription, the most luminous RSGs provide a viable evolutionary pathway to the low luminosity range of the WR population above the \cite{Shenar} stripped star transition between stripped He stars with optically thin winds and WR stars with optically thick winds. Furthermore, our study is restricted to $M < 40\,{\rm M}_\odot$\footnote{Note that the $40\,{\rm M}_\odot$ model appears to enter the forbidden region to the right of the HD limit. We attribute this to insufficient mass loss during the preceding hot-star evolution. When the recently developed hydrodynamically consistent mass-loss rates of \cite{S26b} are adopted, \cite{SV26} find that stars in this mass range turn around near the slanted part of the HD limit.}, and we do not consider the most luminous WR stars, which may form from higher initial masses, potentially through a luminous blue variable (LBV) phase, and/or very massive stars (VMS). 

At the lower mass end, mass-loss rates are insufficient to fully strip the H envelope. The $15\,{\rm M}_\odot$ model fails to form a WR star, and instead reinflates 
and terminates in the YSG region of the HRD while retaining only a relatively low-mass H rich envelope. Its final properties and HRD location are comparable to observed Type IIb supernova (SN) progenitors \citep[e.g.][]{Bersten2012,2016TpyeIIb,2015Smartt}. Therefore, $L/M$ dependent RSG superwinds may represent a single-star evolutionary channel towards producing partially stripped, Type IIb-like pre-SN configurations \citep{2016MNRAS.455..112G}.

Given the nature of mass-loss and the final masses reported in Table~\ref{table} (Appendix~\ref{app:C}), we consider whether the \citetalias{VS23} prescription can reproduce the intermediate mass stripped He stars of \cite{Gotberg}. While the resulting stripped He star masses could, in principle, be produced through single-star mass-loss, the luminosities do not comply (Fig.~\ref{fig:fullgrid}). 
This discrepancy suggests that this population of intermediate-mass stripped stars indeed most likely forms through binary evolution rather than solely single star mass-loss. 

\begin{figure}
    \centering
    \includegraphics[width=1.0\linewidth, height=0.75 \linewidth]{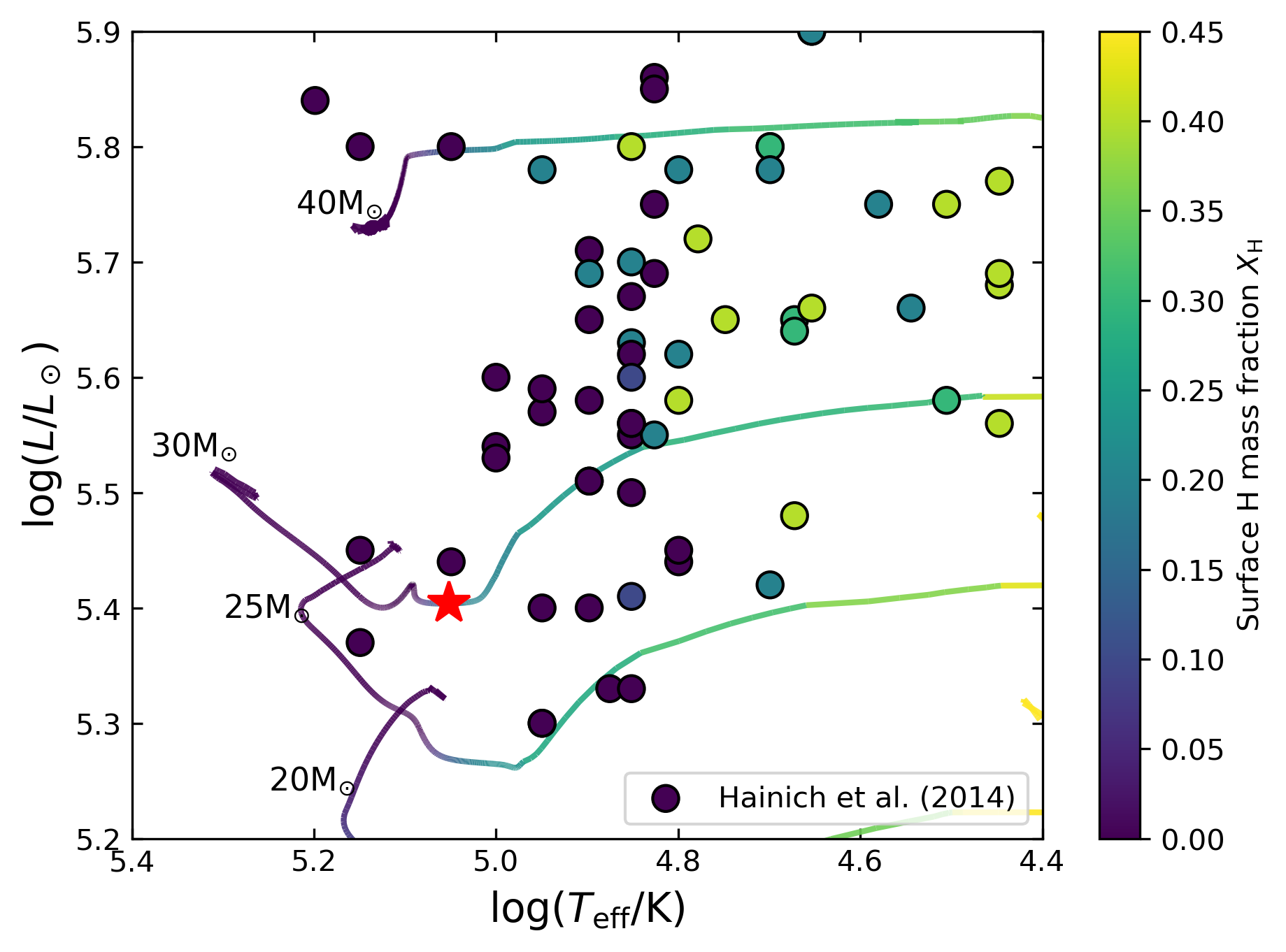}
    \caption{HRD showing the evolutionary tracks of our models with initial masses between $20\,{\rm M}_\odot$ and $40 \,{\rm M}_\odot$ overlaid with the WN population of \citetalias{hainich}. The colour at any point is mapped to the surface H mass fraction given by the colourbar. The red star indicates where we took values from to create a model spectrum (Fig.~\ref{fig:Spectra}). 
    }
    \label{fig:massfracH}
\end{figure}

Stars with initial masses $\geq 20\,{\rm M}_\odot$ shed their H envelopes, resulting in final He surface mass fractions $X_{\rm He}\approx 0.99$ and final nitrogen (N) surface mass fractions $X_{\rm N}\approx 6.4\times 10^{-3}$. This is demonstrated in Fig. \ref{fig:massfracH}, where we show the depletion of surface H due to envelope stripping. In Fig~\ref{fig:massfracH}, we also show the surface H fraction of the observed WN stars from \citetalias{hainich}, which are in qualitative agreement with our models. Combined with the aligned HRD locations discussed earlier, these surface compositions provide additional evidence that the \citetalias{VS23} mass-loss prescription can produce stripped stars with properties consistent with the low-luminosity WN population. We note that the empirical template grids of \citetalias{hainich} adopt a fixed $X_{\rm N} = 4\times 10^{-3}$, somewhat lower than what our models predict.


\begin{figure}
    \centering
    \includegraphics[width=1.0\linewidth]{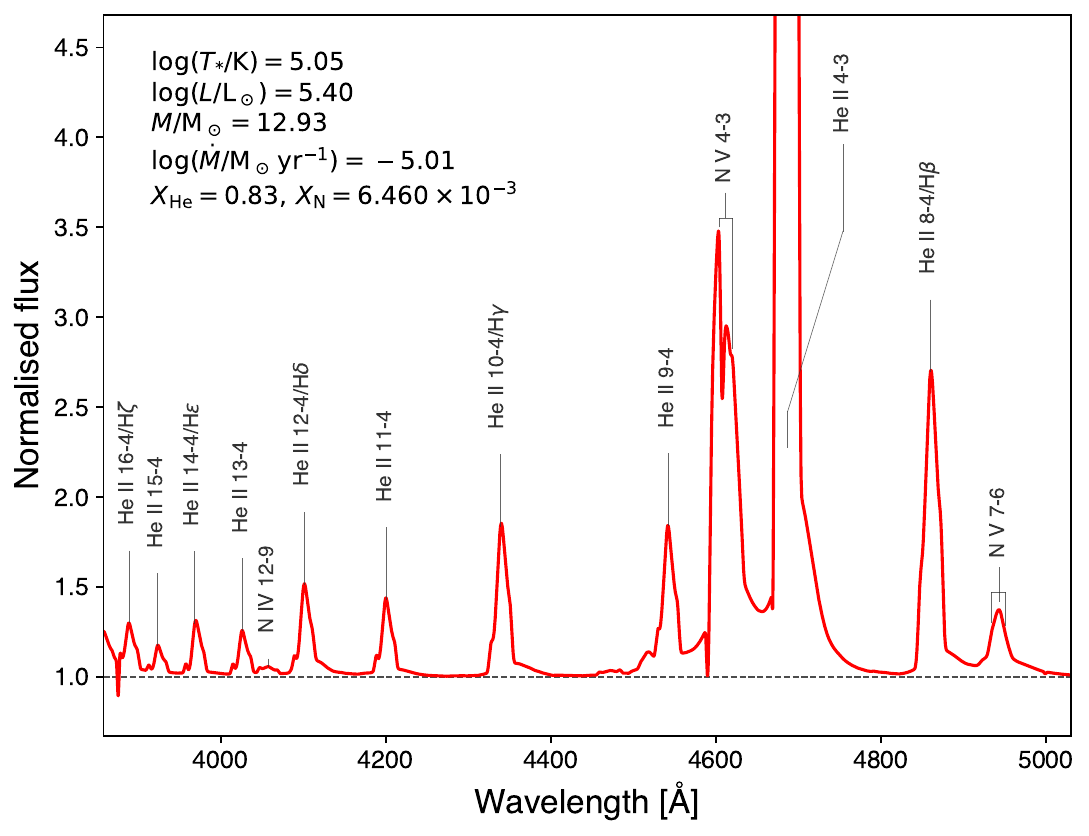}
    \caption{Synthetic PoWR spectrum of the initially $30\, {\rm M}_\odot$ model at the point of the red star in Fig.~\ref{fig:massfracH}. 
    }
    \label{fig:Spectra}
\end{figure}

We further investigate whether WN stars are formed or not by generating a synthetic spectrum of the $30\, {\rm M}_\odot$ initial mass star at the point in its evolution indicated by the red star in Fig.~\ref{fig:massfracH} (details of which are provided in Appendix \ref{app:spectrum}). Our model spectrum, generated with the PoWR code and shown in Fig. \ref{fig:Spectra}, qualitatively captures the characteristic emission features of WN stars \citepalias{hainich}, suggesting that the adopted stellar parameters and surface abundances from our evolutionary models are broadly consistent with those inferred from observed WN stars. 


\begin{figure}
    \centering
    \includegraphics[width=1.0\linewidth, height=0.75 \linewidth]{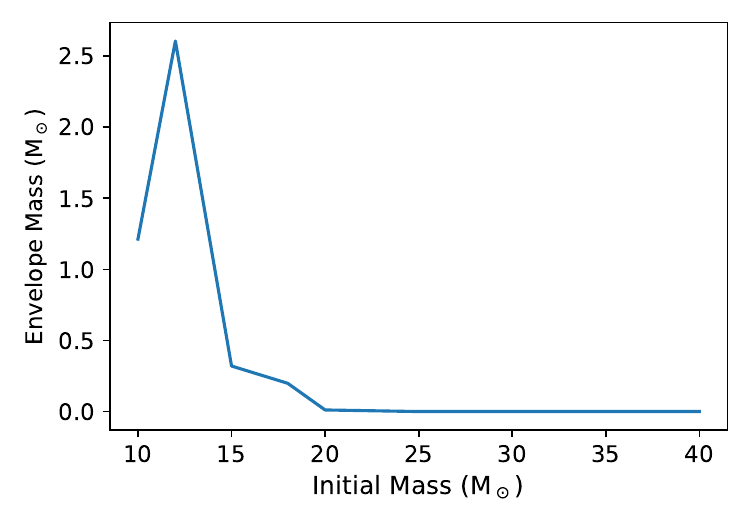}
    \caption{Envelope mass at the end of core He burning (where $Y_c < 10^{-3}$) as a function of the initial mass.}
    \label{fig:m_env}
\end{figure}
The general behaviour of our models is summarised in Fig.~\ref{fig:m_env}, where we plot the remaining envelope mass at the end of core He burning as a function of initial mass. For stars with initial masses of $M>20\,{\rm M}_\odot$, the \citetalias{VS23} winds can strip the H envelope and produce stripped stars consistent with the observed WN population of \citetalias{hainich}. Stars within the range of $15\,{\rm M}_\odot < M< 20\,{\rm M}_\odot$ do not fully strip and instead undergo partial envelope stripping followed by re-expansion.


\section{Conclusions}

In this work, we investigated the impact of RSG mass-loss by comparing two empirical prescriptions: the \citetalias{deJager88} and the more recent \citetalias{VS23} rates, for initial masses between 10$\,{\rm M}_\odot$ and 40$\,{\rm M}_\odot$ at LMC metallicity. We find that adopting the  \citetalias{VS23} prescription allows RSGs, particularly at higher initial masses, to evolve into WR stars through single-star mass-loss alone. This contrasts with models that primarily attribute WR formation in this mass range to binary evolution \citep{Pauli}.

Our results show that the \citetalias{VS23} mass-loss prescription produces partially stripped stars in the intermediate initial-mass range whose luminosities, effective temperatures, and residual H envelope masses are comparable to those inferred for observed Type IIb SNe progenitors. This indicates that enhanced RSG mass-loss can provide a viable single-star evolutionary channel for the formation of Type IIb SN progenitors.

We also find that RSG mass-loss alone is insufficient to reproduce the population of stripped He stars of \cite{Gotberg}, supporting their conclusion that such objects are more likely to form via binary interactions. This implies that, while enhanced RSG winds may contribute to the population of Type Ib/c SN progenitors, particularly at higher initial masses, binary evolution likely remains a necessary channel for producing at least some of the stripped-envelope SN Type Ibc population.

Overall, our results highlight the important role of enhanced RSG mass-loss prescriptions in enabling single star pathways to WR formation and Type IIb SNe progenitors at intermediate to higher masses, while reinforcing the necessity of binary evolution to account for the full observed population of stripped-envelope SN progenitors.

\bibliography{cite}
\bibliographystyle{aa}

\begin{appendix}
    \nolinenumbers


\section{Analytical Mass-Loss Model and Parameter Study}
\label{app:analytical_model}

Out of the luminosity, effective temperature, and mass scaling in RSG mass-loss prescriptions, the luminosity and temperature remain roughly constant during the RSG phase. The mass-loss rate can therefore be approximated as depending only on the current stellar mass, with the other terms absorbed into a renormalised constant, yielding
\begin{equation}
    \log ( \dot{M} )
    =
    \log ( \dot{M}_0 )
    -
    f_m \log \left( \frac{M}{M_{\rm RSG,0}} \right),
    \label{eqn:A1}
\end{equation}
where $\dot{M}_0$ and $M_{\rm RSG,0}$ are the mass-loss rate and stellar mass at the onset of the RSG phase. It is worth briefly noting the physical processes that set the mass-loss rate at the onset of the RSG phase. The core mass at the end of the MS primarily sets the RSG luminosity, and therefore its onset mass-loss rate. The core mass itself is sensitive to the mixing above the convective-radiative boundary, and in particular to the efficiency with which material overshoots this formal boundary. Higher overshooting leads to a larger core mass, which in turn produces a higher RSG luminosity and therefore a higher onset mass-loss rate. Rotation to can impact the luminosity at the end of the MS and therefore the onset RSG mass-loss. 

Exponentiating and identifying $\dot{M} = -{\rm d}M/{\rm d}t$ gives
\begin{equation}
\frac{{\rm d}M}{{\rm d}t}
=
-\dot{M}_0 \left(\frac{M_{\rm RSG,0}}{M}\right)^{f_m}.
\label{eqn:A2}
\end{equation}
Separation of variables and integrating from the beginning of the RSG phase until a certain time $t$, we get
\begin{equation}
    M(t)
    =
    \left[
    M_{\rm RSG,0}^{1+f_m}
    -
    (1+f_m)\dot{M}_0 M_{\rm RSG,0}^{f_m} t
    \right]^{\frac{1}{1+f_m}}.
    \label{eqn:A3}
\end{equation}
At LMC metallicity, the absolute mass-loss rates of typical MS OB stars corresponding to the initial masses considered here are sufficiently low that the total mass lost prior to the RSG phase is small compared to the initial stellar mass. It is therefore a good approximation to set $M_{\rm RSG,0}\approx M_0$, the initial mass. Eq.~\ref{eqn:A3} then simplifies  to
\begin{equation}
    M(t)
    =
    M_0
    \left[
    1
    -
    (1+f_m)\frac{\dot{M}_0}{M_0} t
    \right]^{\frac{1}{1+f_m}}.
    \label{eqn:A4}
\end{equation}


Equation~\ref{eqn:A4} highlights that $f_m$ dictates the mathematical strength of the feedback between a decreasing stellar mass and an increasing mass-loss rate. While this feedback loop can theoretically accelerate envelope depletion, its overall impact depends heavily on the chosen parameter space.

\begin{figure*}
    \centering
    \includegraphics[width=1.0\linewidth]{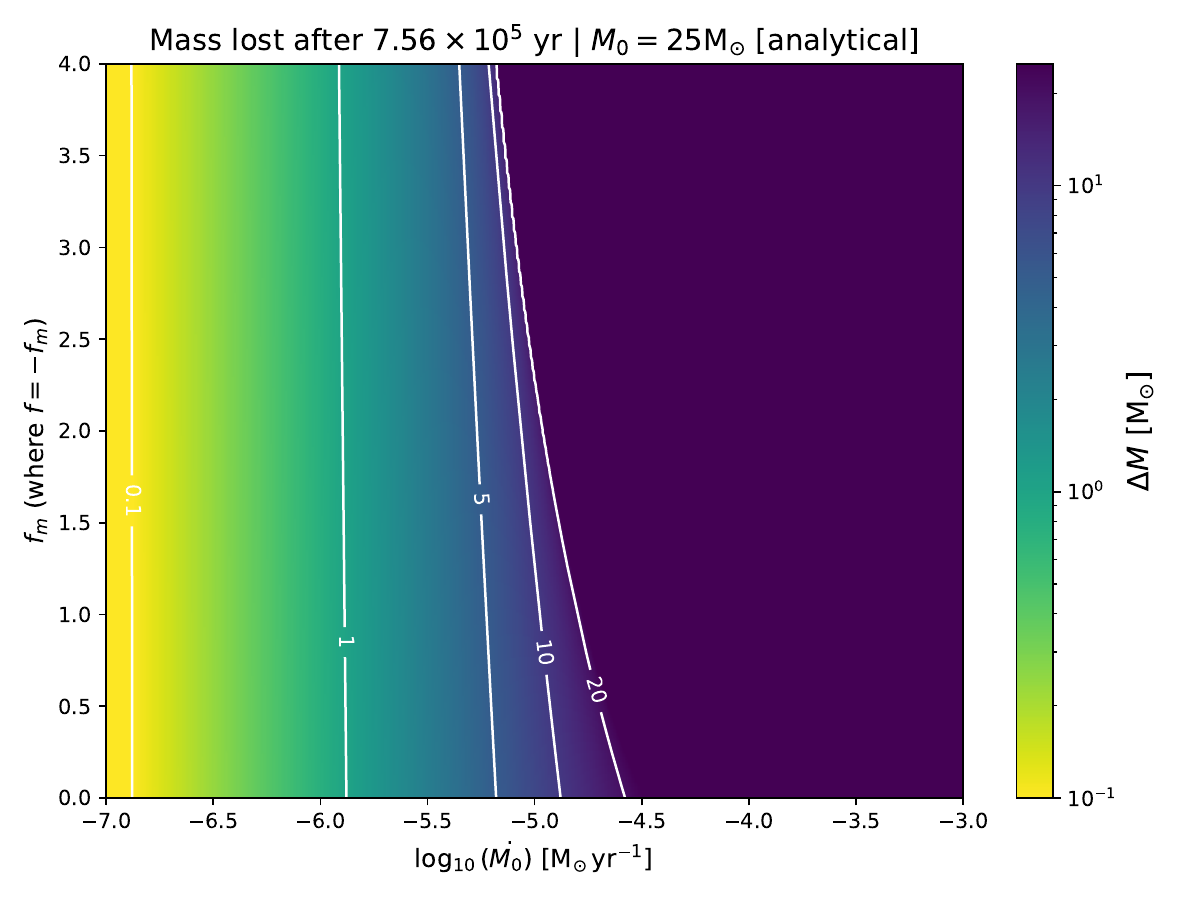}
    \caption{Dependence of total mass-loss on the mass-loss exponent $f_m$ and baseline mass-loss rate $\dot{M}_0$, evaluated within the analytic framework for a representative $25\,{\rm M}_\odot$ progenitor.}
    \label{fig:appendix_1}
\end{figure*}

In Fig.~\ref{fig:appendix_1}, we solve the Eqn.~\ref{eqn:A4} for different values of $f_m$ and $\dot{M}_0$ for a representative $25\,{\rm M}_\odot$ star over its typical core-He burning lifetime of 75,600 years. The figure illustrates the resulting total mass-loss $\Delta M$ as a function of $f_m$ and $\dot{M}_0$, allowing direct comparison of different mass-loss prescriptions within a controlled analytic framework. The analytical model is intentionally decoupled from additional stellar-structure effects, including luminosity evolution, rotation, and convective overshooting, all of which are absorbed into the single free parameter $\dot{M}_0$. The two prescriptions considered in this work correspond to $f_m = 0$ for \citetalias{deJager88}, which has no mass dependence, and $f_m = 3.99$ for \citetalias{VS23}. 

As seen in Fig.~\ref{fig:appendix_1}, at low values of $f_m$ the relevant $5-20\,{\rm M}_\odot$ mass contours are widely spaced, indicating that $\Delta M$ is primarily driven by the baseline mass-loss rate $\dot{M}_0$ with little sensitivity to the mass scaling. At high values of $f_m$, the contours narrow significantly, reflecting a stronger feedback between mass-loss and decreasing stellar mass, and a greater tendency for a runaway stripping of the envelope. This distinction between when the $f_m$ effect begins to dominate has evolutionary consequences. For example, consider an initial mass-loss rate of $\dot{M}_0 \sim 10^{-5}\,{\rm M}_\odot\,\mathrm{yr}^{-1}$ at the onset of RSG phase. If there is no mass feedback as in the \citetalias{deJager88} prescription ($f_m = 0$), then the total envelope stripped is only of order $\sim\!5\,{\rm M}_\odot$ which would not fully strip the envelope. However, if there is a mass feedback similar to the \citetalias{VS23} prescription ($f_m = 3.99$), then the mass-loss runaway is capable of completely stripping the envelope through its strong mass-loss feedback.

\section{Details about Models}
\subsection{MESA}\label{app:MESA_models}
With a metallicity $Z = 0.1$, giving a He mass fraction $Y$ as, $Y =0.24 +2Z$, and H mass fraction $X$ as, $X =1-Y-Z$. Thus, our adopted initial composition is $(X,Y,Z) = (0.73, 0.26, 0.01)$.

Convection was treated using standard mixing length theory (MLT) with a mixing length parameter $\alpha_{\rm{MLT}}=1.5$. We adopted the Ledoux criterion for convective stability, to account for composition gradients in determining convective boundaries. Models were evolved with both MLT++ off and on. In regions that are unstable to semiconvection, mixing and the temperature gradient are calculated following the time-dependent diffusion scheme of \cite{Langer} with a semiconvective efficiency parameter $\alpha_{\rm{sc}} = 1.0$. A step overshoot scheme was adopted, with an overshooting parameter, $\alpha_{\rm{ov}} = 0.3$ for H burning cores and $\alpha_{\rm{ov}} = 0.1$ for He burning cores. Overshooting was applied at the top of the convective core during core H and He burning.

We used the \texttt{basic.net} network with eight isotopes for the reaction network. Opacities were taken from the OPAL Type \textsc{ii} tables, which account for variations in the mass fractions of metals, mainly of carbon and oxygen mass fractions. The equation of state followed a blend of multiple tables depending on the local temperature, density and composition as described in \citet{Jermyn2023}.

\subsubsection{Wind Prescriptions Beyond RSG Winds}\label{app:Other Winds}
Beyond the RSG winds, we implemented a custom mass-loss prescription to handle hot and stripped stellar phases, bypassing MESA's default wind selection logic. For H rich phases ($5\,\rm{kK}<T_{\rm eff}<95\,\rm{kK}$), we implemented the kink concept based on the Monte Carlo results from~\citet{Vink2011}, which captures the mass-loss enhancement due to multiple scattering in the wind. The mass-loss kink was implemented in MESA in~\citet{S22}, where for low-$L/M$, the OB-star prescription of \cite{Vink2001} was used and for high-$L/M$, the scaling from~\citet{Vink2011} was used. Once the star reaches high $T_{\mathrm{eff}}$ ($>100\,\rm{kK}$), the models switch to the hydrodynamically consistent stripped-star wind prescription of \cite{Sander2020}. Smooth transitions are achieved between these different temperature regimes by linearly interpolating between the mass-loss prescriptions, eliminating numerical discontinuities. The mass-loss rate during the pre-main sequence (defined as $X_c / X_{c,\rm init}< 0.999$) is set to zero.

\subsection{Synthetic model spectrum}
\label{app:spectrum}
For our demonstration in Fig. \ref{fig:Spectra}, we use the Potsdam Wolf Rayet Code (PoWR). PoWR \citep{2002A&A...387..244G, 2003A&A...410..993H} is a non-local thermodynamic equilibrium stellar atmosphere code tailored for modelling hot stars with winds. It solves the equation of radiative transfer in the comoving frame, while also assuming statistical and radiative equilibrium. Once convergence is reached, a synthetic emergent spectrum is generated.

As input parameters to the PoWR model used to compute the spectrum presented in Fig. \ref{fig:Spectra}, we use the parameters of the MESA model at the red star in Fig. \ref{fig:massfracH}. That is, $\log(L/{\rm L}_\odot) = 5.4$, $M=12.93\,{\rm M}_\odot$, and $\log(\dot{M}/{\rm M_\odot\,yr}^{-1})=-5.01$, the latter of which is the mass-loss rate of the model at this current point in its evolution given by the prescription of \cite{Vink2001}. We define $L$ and a stellar effective temperature at an inner boundary where the continuum Rosseland optical depth $\tau_{\rm Ross, cont} = 20$, which we call $T_*$ here. To find $T_*$ to input into PoWR, we calculated the radius point where $\tau_{\rm Ross, cont} = 20$ in the MESA model, and found $T_*$ using the Stefan-Boltzmann relation with $L$.
The elements we considered in our calculation were H (with mass fraction $0.1594$), He ($0.832$), C ($7.208\times 10^{-5}$), N ($6.462\times 10^{-3}$), O ($8.4339\times 10^{-4}$), and the Fe-group elements \citep[$7.59\times 10^{-4}$; for details, see][]{2002A&A...387..244G}.

Other input parameters not defined by MESA concern the wind stratification. In our calculation, we assume a standard $\beta$-type velocity law with $\beta=1$ with a terminal velocity $\varv_\infty = 1000\,$km\,s$^{-1}$, and allow for density inhomogeneities under the microclumping approximation, with a constant density contrast $D_{\rm cl} = 4$ throughout the wind. Finally, we adopt a microturbulent velocity of 100\,km\,s$^{-1}$.

\section{Masses \& Final Surface Abundances}\label{app:C}
\begin{table}
\caption{\label{masses}Physical parameters of our \citetalias{VS23} models.}
\centering
\begin{tabular}{ccccccc}
\hline\hline
$M_\mathrm{init}$ &$M_\mathrm{TAMS}$& $M_\mathrm{fin}$ & $M_\mathrm{env,H}$& $X_{\rm He}$ & $X_{\rm N}$ \\
(${\rm M}_\odot$) & (${\rm M}_\odot$) & (${\rm M}_\odot$)&  (${\rm M}_\odot$)&&($\times 10 ^{-3}$)\\
\hline
10&9.846&3.367&1.21&0.551&5.92\\
12&11.80&4.395& 2.61&0.513&5.34\\
15&14.68&5.816&0.32&0.619&6.45\\
18 &17.49& 6.976 &0.20& 0.725&6.56\\
20& 19.32&7.571 &0.01&0.990 &6.43\\
25&23.68&9.555& 0.00&0.990 &6.40\\
30&27.72& 10.52 & 0.00&0.990& 6.38\\
40& 34.79&17.35& 0.00&0.990 & 6.39\\
\hline
\end{tabular}
\tablefoot{$M_{\mathrm{init}}$: initial mass; $M_{\mathrm{TAMS}}$: terminal age main sequence mass; $M_{\mathrm{fin}}$: final mass, which corresponds to the total stellar mass at the end of the simulation. For models that successfully ran to completion, this represents the pre-SN mass; for the 40~${\rm M}_\odot$ model, it represents the mass at the termination of the run, which occurred post-core He burning ($Y_{\mathrm{c}} \approx 1.38 \times 10^{-4}$). $M_\mathrm{env,H}$: remaining H envelope mass at the end of core He burning; $X_{\rm He}$: final surface He mass fraction; $X_{\rm N}$:  final surface N mass fraction.}
\label{table}
\end{table}
\end{appendix}
\end{document}